\documentclass[10pt]{iopart}
\expandafter\let\csname equation*\endcsname\relax
\expandafter\let\csname endequation*\endcsname\relax 

\usepackage{xstring}

\usepackage[shortcuts, acronyms]{glossaries-extra}
\usepackage{graphicx}
\usepackage{dcolumn}
\usepackage{bm}
\usepackage{hyperref}

\usepackage{wasysym}

\usepackage{siunitx}
\DeclareSIUnit\gauss{G}
\usepackage{tensor}
\usepackage{physics}
\usepackage{mathrsfs}  
\usepackage{dsfont}

\usepackage{cite}

\usepackage{floatrow}
\newfloatcommand{capbtabbox}{table}[][\FBwidth]

\setabbreviationstyle[acronym]{long-short}
\newacronym{rf}{RF}{radiofrequency}
\newacronym{srf}{single-RF}{single radiofrequency}
\newacronym[firstplural = multiple radiofrequencies (multi-RFs)]{mrf}{multi-RF}{multiple-radiofrequency}
\newacronym{bec}{BEC}{Bose-Einstein condensate}
\newacronym{tof}{TOF}{time-of-flight}
\newacronym{rwa}{RWA}{rotating wave approximation}

\newcommand{\twopi}{2\pi\cdot}
\newcommand{\id}{\mathds{1}}
\newcommand{\rb}{$^{87}$Rb}
\newcommand{\ie}{i.\,e.}
\newcommand{\eg}{e.\,g.}
\newcommand{\tp}{t_\text{p}}
\newcommand{\Vp}{V_\text{p}}
\newcommand{\omegap}{\omega_\text{p}}
\newcommand{\omegarf}{\omega_\text{rf}}
\newcommand{\omegaf}{\omega_\text{f}}
\newcommand{\aprobe}{\widetilde{a}_\text{p}}
\newcommand{\afield}{\widetilde{a}}
\newcommand{\np}{N_\text{p}}
\newcommand{\nph}{\hat{N}_\text{p}}
\newcommand{\Vrf}{V_\text{rf}}
\newcommand{\ketprime}[1]{\tensor[]{\ket{#1}}{_1}}

\newcommand{\melprime}[3]{\tensor[_1]{\mel{#1}{#2}{#3}}{_1}}

\newcommand{\Omegap}[1]{\Omega_{#1\text{+}}}
\newcommand{\Omegam}[1]{\Omega_{#1\text{--}}}
\newcommand{\Omegaz}[1]{\Omega_{#1\text{z}}}
\newcommand{\Omegaeff}{\Omega_\text{eff}}
\newcommand{\ki}[1]{n_{#1}}
\newcommand{\subspace}{\mathscr{E}_0}
\newcommand{\deltak}{\kappa}

\renewcommand{\emph}[1]{\textcolor{red}{#1}}

\usepackage{xcolor}

\definecolor{Omega}{rgb}{0.0353, 0.3804, 0.2413}
\definecolor{omegaPOmega}{rgb}{0.102, 0.4824, 0.3294}
\definecolor{omegaMOmega}{rgb}{0.1882, 0.5608, 0.4118}
\definecolor{OmegaH}{rgb}{0.2392, 0.0549, 0.3765}
\definecolor{HomegaPOmega}{rgb}{0.3255, 0.1255, 0.4745}
\definecolor{HomegaMOmega}{rgb}{0.4039, 0.2118, 0.5569}
\definecolor{omegaPOmegaH}{rgb}{0.5059, 0.3255, 0.6431}
\definecolor{omegaMOmegaH}{rgb}{0.6588, 0.5098, 0.7725}

\begin{document}
	
\title[]{Probing multiple-frequency atom-photon interactions with ultracold atoms}

\author{K Luksch$^1$, E Bentine$^1$, A J Barker$^1$, S Sunami$^1$, T L Harte$^{1,2}$, B Yuen$^1$ and C J Foot$^1$}

\address{$^1$ Clarendon Laboratory, University of Oxford, Parks Road, Oxford OX1 3PU, United Kingdom}
\address{$^2$ Cavendish Laboratory, University of Cambridge, J.\ J.\ Thomson Avenue, Cambridge CB3 0HE, United Kingdom}
\ead{christopher.foot@physics.ox.ac.uk}
\vspace{10pt}
\begin{indented}
\item[]\today
\end{indented}

\begin{abstract}
We dress atoms with multiple-radiofrequency fields and investigate the spectrum of transitions driven by an additional probe field. A complete theoretical description of this rich spectrum is presented, in which we find allowed transitions and determine their amplitudes using the resolvent formalism. Experimentally, we observe transitions up to sixth order in the probe field using radiofrequency spectroscopy of Bose-Einstein condensates trapped in single- and multiple-radiofrequency-dressed potentials. We find excellent agreement between theory and experiment, including the prediction and verification of previously unobserved transitions, even in the single-radiofrequency case.
\end{abstract}

% Uncomment for keywords - min 3, max 7 keywords

\vspace{2pc}


\section{Introduction}
Spectroscopy has spurred great progress in our understanding of physical systems, from the quantum-mechanical explanations of the hydrogen spectrum to measurements of the Lamb shift~\cite{Lamb1947}. Precision measurements continue to illuminate the limits of our knowledge~\cite{Antognini2013, Beyer2017}. In turn, the experimental tools developed from spectroscopy have advanced our ability to manipulate the external and internal degrees of freedom of atoms. For instance, controlling the motion of atoms using light has led to laser-cooling and spatial confinement of atomic vapours. Meanwhile, controlling the quantum state has provided essential tools to investigate the fundamental principles of quantum mechanics~\cite{Itano1990, Hosten2016} and is intrinsic to quantum information processing~\cite{Monz2011,Saffman2016,Levine2018}.

The dressed-atom formalism~\cite{Cohen-Tannoudji1992} is an established framework for understanding atom-photon interactions. Applications include laser cooling, cavity quantum electrodynamics, and trapping of cold atoms; the latter encompasses atoms dressed with optical~\cite{Grimm2000}, microwave~\cite{Spreeuw1994} or \ac{rf}~\cite{Zobay2001, Colombe2004, Schumm2005,Garraway2016} radiation to either provide or shape the confinement.  Significant attention has been paid to confining atoms in single-frequency dressing fields. In addition, a weak probe field is often used to drive transitions between dressed states~\cite{Happer1964,Allegrini1971}. 

\glsunset{srf}

There are a number of physical systems in which an atom interacts with multiple-frequency radiation.
In the field of non-linear optics, examples such as four-wave mixing and electromagnetically induced transparency have been studied extensively~\cite{Fleischhauer2005,Bloembergen1996}. Spectroscopic signals can be resonantly enhanced by the use of multiple frequencies, such as in stimulated Raman spectroscopy and coherent anti-Stokes Raman spectroscopy~\cite{Prince2017,Evans2008}.
Optical dipole traps composed of multiple frequencies can be used to confine quantum gases in superlattices~\cite{Sebby-Strabley2006,Jo2012} or species-selective potentials~\cite{LeBlanc2007}. In these cases, the radiation is far from resonance and the perturbation arising from each frequency component can be treated independently.
Using this approach to describe atoms dressed with multiple radiofrequencies (multi-RFs) as in~\cite{Courteille2006} is inaccurate when coherent processes are important, such as for separations between frequency components comparable to or smaller than the Rabi frequencies.

\glsunset{mrf}

We have recently addressed this issue, and demonstrated the use of a \ac{mrf} dressing field to confine atoms in a double-well potential~\cite{Harte2018}. Atoms are trapped in two parallel sheets, the spacing between which can be made sufficiently small to realize matter-wave interferometry~\cite{Bentine2018}. These potentials are smooth, with a tunable geometry and the capacity to influence the dimensionality of the trapped gas~\cite{Harte2018,Merloti2013}. 

A large number of possible transitions arise for atoms dressed with multiple fields, which renders these potentials vulnerable to atom loss caused by \ac{rf} noise. The versatility of these potentials prompted an investigation into the spectrum of transitions as a means to investigate susceptibility to noise. In turn, this has resulted in the theoretical framework presented here, which is applicable to a wide range of systems dressed by multiple frequencies. Transitions in atoms in \ac{rf}-dressed potentials have been calculated in previous work, though restricted to a single dressing frequency~\cite{GarridoAlzar2006,Easwaran2010,Perrin2017}. Moreover, only first-order transitions were considered in~\cite{GarridoAlzar2006,Easwaran2010} and higher-order transitions were calculated only for selected polarizations in~\cite{Perrin2017}.

In this paper, we study the spectra of atoms dressed by single- and multiple-frequency fields. We present experimental spectra, in which Bose-Einstein condensates (BECs) of \rb{} atoms are dressed and probed by \ac{rf} fields. For a single-frequency dressing field, accounting for all polarizations and higher-order transitions results in a spectrum that goes beyond the well-known Autler-Townes splitting~\cite{Autler1955}. For certain polarizations of the probe field, we identify resonances that have not been observed or predicted previously. This is further generalized by including multiple dressing frequencies, revealing a rich spectrum of resonant transitions that we fully characterize spectroscopically.

\glsunset{bec}

We start by detailing a theoretical framework that fully characterizes transitions excited by a probe field when an atom is dressed by a multi-frequency field (\sref{sec:theory}). In \sref{sec:experiment} the experiment is described. In \sref{sec:theoryResults} we demonstrate the versatility of the theoretical framework by performing detailed calculations for transitions of any order in the probe field and compare these to experimental results in \sref{sec:comparison}.

\section{Theory of transitions in the dressed-atom picture}
\label{sec:theory}
We consider an atom in a static magnetic field, dressed by a coherent radiation field with one or more frequency components. We calculate resonant frequencies and coupling strengths for transitions driven by a coherent probe field between the dressed eigenstates. Dressing the atoms leads to a ladder of energy levels giving rise to a spectrum that is considerably more complex than that of the bare atoms. In the following, $g_F < 0$ and $F = 1$ are chosen for the examples, for consistency with the \rb{} $F = 1$ hyperfine manifold which we investigate experimentally.

\begin{figure}[h]
	\includegraphics[width = \textwidth]{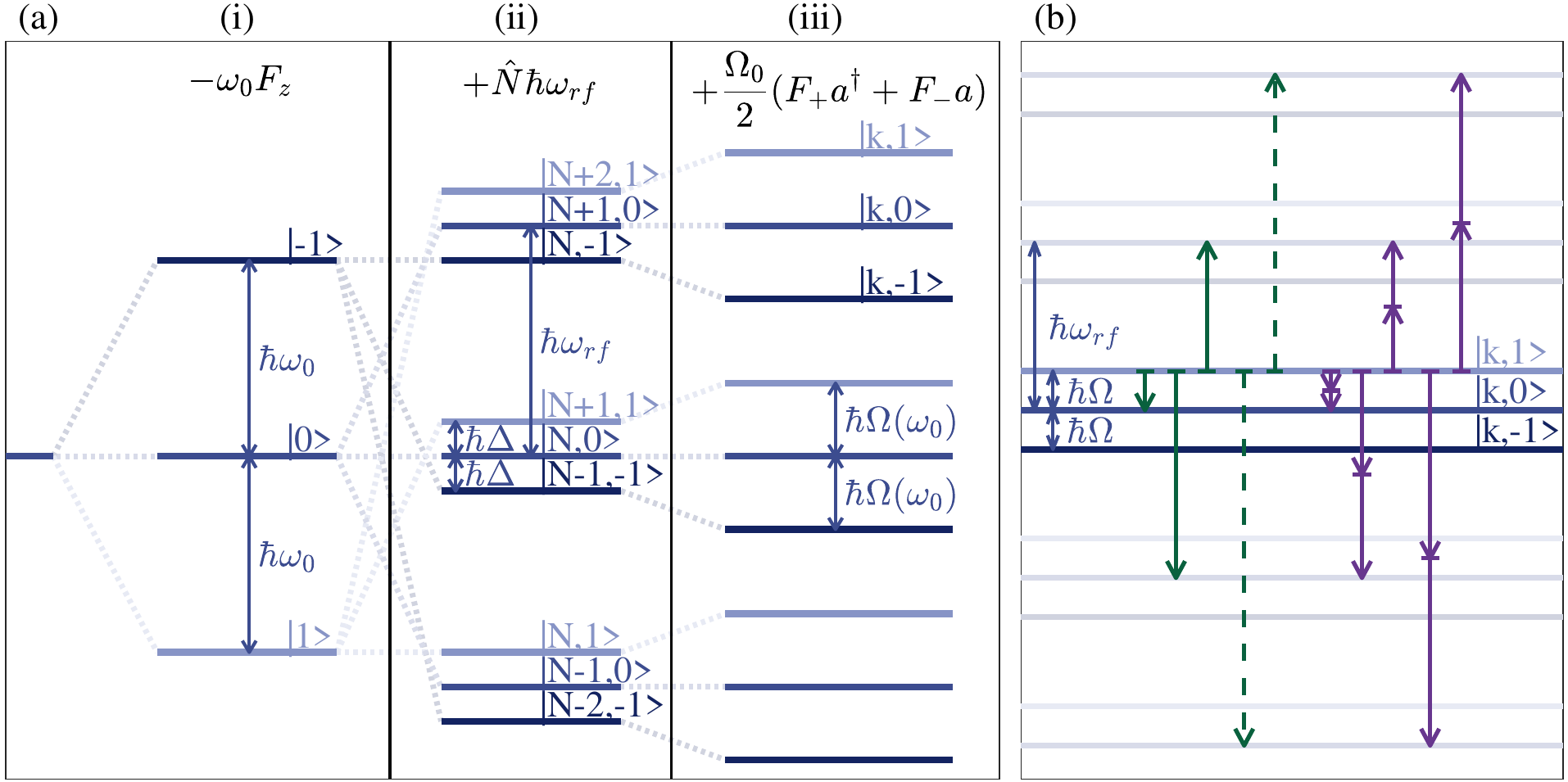}
	\caption{\label{fig:states} (a) Eigensystems of various components of the Hamiltonian under consideration, for an atom with $g_F < 0$ and $F = 1$. For simplicity, these are shown for a single-frequency, circularly-polarized dressing field at frequency $\omegarf$. (i) The Zeeman effect of a static magnetic field lifts the degeneracy of the three energy levels. (ii) The addition of the dressing field results in a ladder of energy levels (probe field omitted for clarity). (iii) Taking into account the interaction between atoms and dressing field gives the dressed eigenstates. $\Omega(\omega_0)$ depends on $\omega_0$ and is equal to the Rabi frequency on resonance  $\Omega_0$. (b) Illustration of some first- (green) and second-order (purple) transitions of the dressed atoms. All transitions which exist for circular dressing are shown. Dashed lines indicate transitions which arise from terms beyond the \ac{rwa} of the dressing field.}
\end{figure}

\Fref{fig:states} illustrates the bare and dressed states and the transitions which can be driven resonantly by the probe field. We calculate transition frequencies and amplitudes for the Hamiltonian
\begin{eqnarray} \label{eq:HandH0}
H = H_0 + \Vrf + \Vp, \quad \text{with} \quad
H_0 =  -\omega_0 F_z + \hat{N} \hbar \omegaf + \nph\hbar\omegap.
\end{eqnarray}
The interaction-free part of the Hamiltonian, $H_0$, is the sum of the atomic energy $\hbar \omega_0 = |g_F|\mu_B B$ of an atom in a static field, and the field energies $\hat{N} \hbar \omegaf$ and $\nph \hbar \omegap$ of the time-dependent dressing and probe fields. The atom remains in the electronic ground state at all times, hence spontaneous emission is negligible. We consider dressing fields with angular frequencies $\ki{q}\omegaf$, which are integer multiples of a common fundamental frequency $\omegaf$, and the probe field with angular frequency $\omegap$. The fundamental frequency $\omegaf$ is defined such that the $\ki{q}$ are coprime integers.

$\Vrf$ and $\Vp$ describe the interaction of the atom with the dressing fields and with the probe field, respectively. The dressing field $\Vrf$ is turned on adiabatically, such that as the atoms are dressed by the field, their states become eigenstates of $H_1 = H_0 + \Vrf$ (\fref{fig:states}~(a.iii)). The probe field, which is pulsed on non-adiabatically, drives transitions between these dressed states (\fref{fig:states}~(b)). 

To calculate the strength of these transitions we must find eigenstates of $H_1$ and then calculate matrix elements of $\Vp$ between them. Extending the single-frequency dressed-atom picture, a natural choice of basis is  to represent the coherent state of the dressing field using tensor products of Fock states.
For multiple dressing fields, however, this basis is degenerate, leading to complications during diagonalization.
%
% V1.0
%Instead, we work within a subspace of the Hilbert space that is sufficient to describe the coherent states of the problem. This subspace is spanned by the basis $\lbrace \ket{N} \rbrace$, where each basis state $\ket{N}$ is a superposition of the Fock states with energy $\hat{N}\hbar \omegaf \ket{N} = N\hbar\omegaf \ket{N}$.
%Further details can be found in \ref{sec:basis} with more detailed mathematical proofs in~\cite{Ben} and another example of an application in~\cite{Yuen2018}. 
% 
%
Instead, we work within a subspace of the Hilbert space that is spanned by the non-degenerate orthonormal set $\lbrace \ket{N} \rbrace$, where each state $\ket{N}$ is a superposition of Fock states with energy $\hat{N}\hbar \omegaf \ket{N} = N\hbar\omegaf \ket{N}$. 
This set of non-degenerate states can be used as a basis to describe our system and to calculate transitions as explained in \ref{sec:basis}.
More detailed mathematical proofs can be found in~\cite{Ben} and another example of an application in~\cite{Yuen2018}.
The probe field is treated separately in the standard Fock basis, with the field energy $\nph\hbar\omegap\ket{\np} = \np\hbar\omegap\ket{\np}$.

The eigenstates of $H_0$ are $\ket{N,\np,m_F}_0$, where $\hbar m_F$ is the component of the atom's spin projected along $z$. The eigenenergies $\hbar(N\omegaf+\np\omegap-m_F\omega_0)$ form a ladder as shown in \fref{fig:states}~(a.ii). The interactions of the dressing and probe fields with the atom are
\begin{eqnarray}
\label{eq:Vs}
\fl
\Vrf =  \sum_q \bigl[\frac{1}{2}(\Omegap{q} F_+ \afield_q + \Omegap{q}^*F_- \afield_q^\dagger + \Omegam{q} F_- \afield_q + \Omegam{q}^*F_+ \afield_q^\dagger ) + (\Omegaz{q}\afield_q + \Omegaz{q}^*\afield_q^\dagger) F_z\bigr],\\
\label{eq:Vp}
\fl
\Vp =  \frac{1}{2}(\Omegap{}\aprobe F_++\Omegap{}^*\aprobe^\dagger F_-+\Omegam{}\aprobe F_-+\Omegam{}^*\aprobe^\dagger F_+) + (\Omegaz{}\aprobe + \Omegaz{}^*\aprobe^\dagger) F_z,
\end{eqnarray}
with $F_\pm = F_x\pm iF_y$, where $F_x,F_y,F_z$ are the spin projection operators. The Rabi frequencies $\Omega_{q,\pm,z},\Omega_{\pm,z}$ are complex numbers and thus account for the relative phase between the multiple fields. 
The creation and annihilation operators commute to a very good approximation, since we consider coherent fields with large mean photon numbers. We can therefore define normalized raising and lowering operators $\afield_q^\dagger, \afield_q, \aprobe^\dagger, \aprobe$, which act on the corresponding basis states such that 
\begin{eqnarray}
\label{eq:normaloperators}
\afield_q\ket{N} = \ket{N-\ki{q}},\ \afield_q^\dagger\ket{N} = \ket{N+\ki{q}},\\ 
\aprobe^\dagger\ket{\np} = \ket{\np+1},\ \aprobe\ket{\np} = \ket{\np-1}.
\end{eqnarray}
We group the eigenstates of $H_1$ by their energy, labelled by the index $k$, and denote them by $\ketprime{k,\np,m}$ with eigenenergies
\begin{equation}
\label{eq:eigenenergies}
H_1\ketprime{k,\np,m} = \hbar\left(k\omegaf+\np\omegap+m\Omega(\omega_0)\right)\ketprime{k,\np,m}.
\end{equation} 
The quantum number $m$ takes values between $-F$ and $F$ in integer steps, and $\hbar\Omega(\omega_0)$ corresponds to the energy difference between neighbouring states of equal $k$, as shown in \fref{fig:states}~(a.iii). Additionally, we define $\Omega(\omega_0)\leq \omegaf/2$ for integer values of $F$ and $\Omega(\omega_0) \leq \omegaf$ for half-integer values of $F$. Note that these are locally-defined quantum numbers which differ from the commonly used case (see \ref{sec:manifolds}). The terms in \eref{eq:Vs} and \eref{eq:Vp} with Rabi frequencies $\Omegam{q},\Omegam{}$ couple states resonantly for $g_F < 0$. The off-resonant terms with Rabi frequencies $\Omegap{q},\Omegap{}$ are neglected under the \ac{rwa}.

Having described the Hamiltonian of the dressed-atom system, we now examine which transitions are allowed between the eigenstates of $H_1$. We call the order of a transition the total number of probe photons created and/or annihilated in driving the system from the initial to the final eigenstate, as illustrated in \fref{fig:states}~(b). Note that an arbitrary number of photons of the dressing field can be involved, resulting in an unlimited number of transitions compared to very few in the case of undressed states.

For a transition to occur, two conditions need to be met: (i) the probe interaction couples initial and final states and (ii) the probe frequency is resonant, such that energy is conserved. For first-order transitions from initial state $\ketprime{k,\np,m}$ to final state $\ketprime{k',\np',m'}$ with $m'\neq m$, these conditions are expressed as:
\begin{eqnarray}
\label{eq:firstorder}
\frac{\Omegaeff}{2}\mel{m'}{F_{\text{sgn}(m'-m)}^{|m'-m|}}{m} & = \melprime{k',\np',m'}{\Vp}{k,\np,m}\neq 0, \quad \text{and}\\
\label{eq:energyconservation}
k'\omegaf + \np'\omegap + m'\Omega & = k\omegaf + \np\omegap + m\Omega,
\end{eqnarray}
where we have defined an effective Rabi frequency $\Omegaeff$, and $F_{\text{sgn}(m'-m)}^{|m'-m|}$ denotes the application of the spin raising or lowering operator $|m'-m|$ times to connect states $\ket{m}$ and $\ket{m'}$. 

Higher-order transitions arise when the path taken between the initial and final state includes a number of off-resonant intermediate states. The resultant transition amplitude therefore depends on the amplitudes of these individual paths, which may interfere. To determine the frequencies of transitions and calculate their strengths, we use the resolvent formalism~\cite{Cohen-Tannoudji1992,Allegrini1971}. This gives an effective Hamiltonian which can be used to calculate transitions of any order, as explained in \ref{sec:resolvent}. These higher-order transitions are found by replacing \eref{eq:firstorder} with the following
\begin{equation}
\label{eq:norder}
\frac{\Omegaeff}{2}\mel{m'}{F_{\text{sgn}(m'-m)}^{|m'-m|}}{m} = \melprime{k',\np',m'}{\Vp\frac{Q}{E_0-H_1}\Vp\cdots\Vp\frac{Q}{E_0-H_1}\Vp}{k,\np,m}\neq 0.
\end{equation}
$E_0$ is the energy of the initial state and $Q$ a projection operator onto all states with energy not close to $E_0$. Following the explanation above, \eref{eq:norder} can be interpreted as follows: the initial state interacts with the probe field, followed by a free evolution under $H_1$, followed by another interaction and so forth, connecting initial and final states by $i$ interaction terms for $i^{th}$ order. The condition for energy conservation expressed in \eref{eq:energyconservation} must also be fulfilled for higher orders.

\section{Spectroscopy of radiofrequency-dressed potentials}
\label{sec:experiment}
\begin{figure}[h]
	\includegraphics[width = \textwidth]{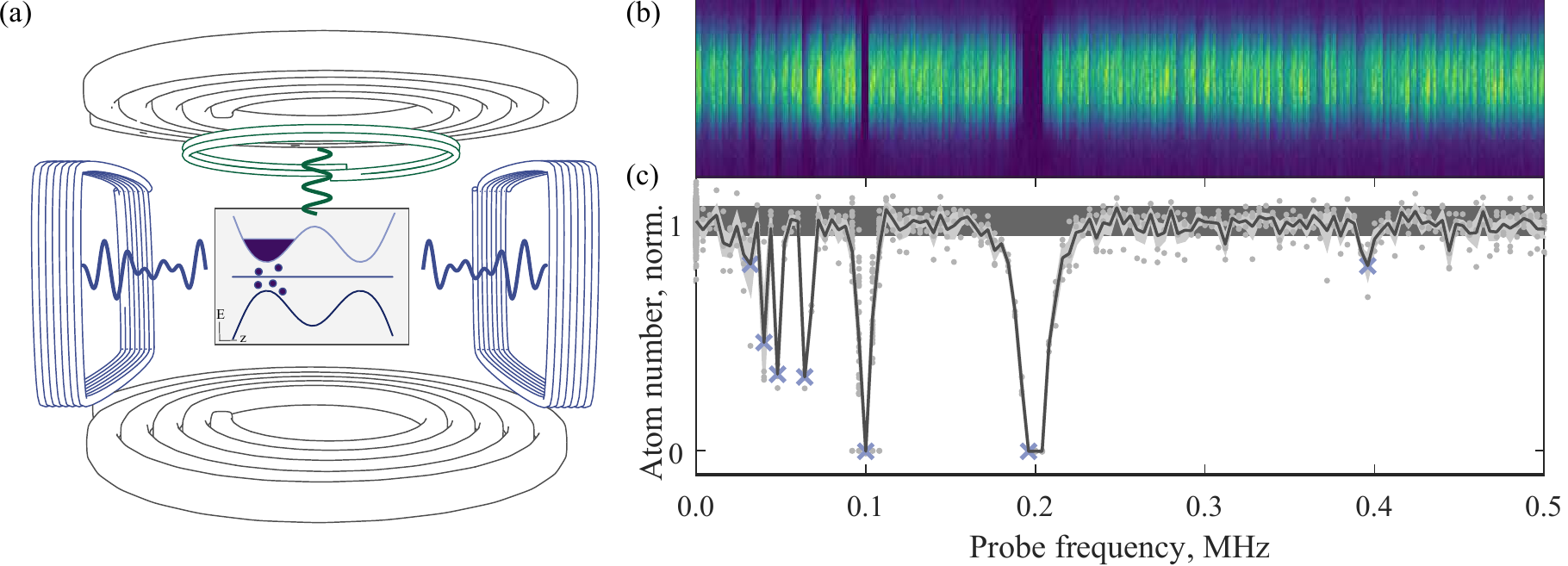}
	\caption{\label{fig:experiment} (a) The inset shows atoms (purple) trapped in a potential (blue) by dressing a static magnetic quadrupole field with multiple RFs~\cite{Harte2018}. The energies of three dressed eigenstates are indicated, with atoms trapped in the upper state. Atoms are lost from the trap if the probe field drives them to an untrapped state. Coils for the generation of the static quadrupole field (grey, top and bottom), dressing fields (blue, left and right), and evaporative cooling (green, top) are shown. (b) Slices of absorption images and (c) normalized atom number versus probe frequency for atoms trapped in a linearly-polarized dressing field of a single frequency. The grey points indicate measured atom numbers (normalized), and the black line shows the average at each probe frequency. The light grey area indicates the standard error of the mean. The dark grey strip indicates the range of atom numbers within one standard deviation of the background distribution. Resonances identified from the data using the Kolmogorov-Smirnov test are marked with crosses.}
\end{figure}

Having described the theory of transitions in \ac{rf}-dressed potentials, we now present the details of the experimental work. Our apparatus traps a cloud of ultracold \rb{} atoms in a potential created by the spatial dependence of the dressed eigenenergies in a static quadrupole field, dressed by one or more RFs~\cite{Harte2018,Zobay2001}, as shown in \fref{fig:experiment}~(a). A probe field drives transitions to untrapped states, leading to atom loss, which we measure through absorption imaging of the remaining atoms. The apparatus and experimental sequence is as described previously~\cite{Harte2018}. 

Centimetre-scale coils generate both the static and \ac{rf} fields, with a static quadrupole field $\vec{B} = B'(x\vec{e}_x + y\vec{e}_y - 2z\vec{e}_z)$. The atomic cloud is evaporatively cooled to a \ac{bec} of approximately $2\cdot10^5$ atoms and loaded into either a single or a multiple \ac{rf}-dressed potential~\cite{Harte2018}. In all sequences, atoms are confined in the potential well formed by an \ac{rf}-dressing field with a frequency of \SI{3.6}{\MHz}, and $B' = \SI{155}{\gauss\per\cm}$.

To perform the spectroscopy of \ac{mrf} dressed states, we trap atoms in the state with $m = 1$ in a \ac{mrf}-dressed potential and apply an additional probe field for a duration $\tp = \SI{1.2}{\s}$. On resonance, the probe field drives atoms to untrapped states with $m' = 0, -1$. We observe transitions by measuring the remaining atom number after \ac{tof} expansion using absorption imaging. A collage of sample images is shown in \fref{fig:experiment} (b). The measurement is repeated four times for each probe frequency and the frequency is varied with a spacing of \SI{4}{\kHz} for \ac{srf} and \SI{2}{\kHz} for \ac{mrf} dressing.

For the \ac{srf} case, we use a dressing field with an amplitude of $\Omegam{1} = \twopi\SI{197}{\kHz}$ and $\Omegap{1} = 0$ for a circularly-polarized field or $\Omegap{1} = \Omegam{1}$ for a linearly-polarized field. For the \ac{mrf} case, atoms are irradiated by dressing fields with frequencies of \SIlist[list-units=single]{3.6; 3.8; 4.0}{\MHz}, corresponding to $\omegaf = \SI{0.2}{\MHz}$ and $\ki{1},\ki{2},\ki{3} = 18,19,20$ with Rabi frequencies \SIlist[list-units=single]{80; 69; 99}{\kHz}. These additional dressing field components displace the position of the atoms in the well formed by $\omega_1$ to $\omega_0 = \twopi\SI{3.6383}{\MHz}$, resulting in a resonance at $\Omega(\omega_0) = \twopi\SI{72.5}{\kHz}$~\cite{Harte2018}.

The antenna we use for the spectroscopy pulses is situated above and perpendicular to the coils generating the dressing fields, as shown in \fref{fig:experiment} (a). This results in a field with predominantly linear polarization along the $z$ direction, \ie{} $\Omegaz{}\gg \Omega_\pm$. 
We estimate the Rabi frequency of the applied field at \SI{3.6}{\MHz} as $\Omegaz{} = \twopi\SI{30}{\kHz}$ by dressing trapped atoms with the probe field and measuring the displacement in the horizontal direction~\cite{Sherlock2011,Gildemeister2012}.

The characteristics of the amplifier and impedance of the coil cause the probe amplitude to vary by \SI{37}{\decibel} over the frequency range of interest. Below \SI{0.7}{\MHz}, the amplitude drops by approximately \SI{30}{\decibel}, such that $\Omegaz{} < \SI{1}{\kHz}$. It increases by \SI{7}{\decibel} in the range between \SIlist{0.7; 7.5}{\MHz}, with a self-resonance near \SI{4.7}{\MHz}. We use the maximum probe amplitude available, only reducing it over the range \SIrange{4.3}{5.1}{\MHz} to compensate for the coil resonance. For probe frequencies above \SI{2}{\MHz}, $\Omegaz{}$ is within an order of magnitude of the dressing field amplitudes, such that the assumption of a weak probe field does not hold. This perturbs the resonance frequencies, but does not change the existence of these resonances. We therefore use the maximum amplitude, as our aim is to determine the existence of numerous resonances.

The atom number prepared in each experimental cycle fluctuates randomly by about \SI{10}{\%}, and a dip in measured atom number due to a resonance must be reliably distinguished from this random noise. We determine this background statistical distribution by measuring the atom number (without a probe applied) at random times interleaved with each data series. Resonant frequencies are identified as those for which the mean atom number corresponds to a minimum, and the distribution of measured atom numbers is different from the background distribution, using the Kolmogorov-Smirnov test with a significance level of 0.01.

Acquiring detailed spectra with many repeats and a cycle time of approximately one minute requires continuous operation for days at a time. The atom number prepared in our experiment typically drifts by \SI{30}{\%} over many hours. To reduce systematic effects, measurements and repeats at different frequencies were taken in a random, interspersed order. Furthermore, to remove long-term drifts, we normalize each atom number measurement with respect to the mean atom number in an interval spanning 30 minutes either side of that point, and excluding points with an applied probe RF on or close to resonance as determined by the Kolmogorov-Smirnov test.

We apply this procedure to measure the loss spectrum of atoms trapped in single- and multi-frequency fields. \Fref{fig:experiment} (c) shows a sample section of the spectroscopic measurement for a linearly-polarized \ac{srf} dressing field, with the identified resonances indicated.

\section{Theoretical predictions}
\label{sec:theoryResults}
The general formalism in \sref{sec:theory} is applicable to any system of dressed eigenstates that are subjected to a non-adiabatic perturbation. We calculate transitions for specific examples relating to the experimental work described above, namely \rb{} atoms in the hyperfine ground state ($F = 1$) with $m = 1$, and transitions from this initial state to untrapped states with $m' = 0$ or $m' = -1$.
\subsection{Single radiofrequency -- circular polarization}
\label{sec:theoryCirc}
We first consider the simple example of a circularly-polarized dressing field with a single frequency $\omegarf$, where there are no counter-rotating terms. In this simple case, the interaction with the dressing field reduces to $\Vrf = (\Omega_0/2)\left(F_+\afield^\dagger + F_-\afield\right)$, where we can assume a real Rabi frequency $\Omega_0$. The interaction is confined to one manifold with constant $k_c = N-m_F$. This means $H_1$ can be diagonalized exactly, and its eigenstates are admixtures of states from a single manifold. For our experimental parameters, this definition coincides with that of \eref{eq:eigenenergies} (see \ref{sec:manifolds}).

To facilitate the calculation of matrix elements, we rewrite the probe interaction $\Vp$ (defined in \eref{eq:Vp}) in terms of operators acting on eigenstates of $H_1$ such that
\begin{eqnarray*}
	S_\pm\ketprime{k_c,\np,m} & = \hbar\sqrt{F(F+1)-(m\pm1)m}\ketprime{k_c,\np,m\pm 1},\\ S_z\ketprime{k_c,\np,m} & = \hbar m\ketprime{k_c,\np,m},\\ 
	b^\dagger\ketprime{k_c,\np,m} & = \ketprime{k_c+1,\np,m}, \\
	b\ketprime{k_c,\np,m} & = \ketprime{k_c-1,\np,m}.
\end{eqnarray*}
This results in
\begin{eqnarray}
\nonumber
\fl
\Vp = \frac{1}{2\Omega_c}\Bigl[ \left( \Omegaz{}\aprobe^\dagger + \Omegaz{}^*\aprobe \right)\left(\Omega_0\left(S_+ + S_-\right) + 2\Delta S_z  \right) + \\
\nonumber
\frac{1}{2}\Bigl( \Omegap{}\aprobe b\left((\Omega_c - \Delta) S_- +2\Omega_0S_z - (\Omega_c+\Delta)S_+ \right) + 
h.c. \Bigr) + \\
\frac{1}{2} \Bigl(\Omegam{} \aprobe b^\dagger\left((\Omega_c - \Delta) S_+ +2\Omega_0S_z - (\Omega_c + \Delta) S_- \right) + 
h.c. \Bigr)
\Bigr],
\label{eq:Vpt}
\end{eqnarray}
with the angular frequency detuning $\Delta = \omegarf - \omega_0$. For this simplified case, the generalized Rabi frequency of the dressing field is $\Omega_c^2 = \Omega_0^2+\Delta^2$.

We determine the probe frequencies for resonant transitions, and calculate their corresponding strength, by solving \eref{eq:energyconservation} and \eref{eq:norder}. Transitions of order $i$ exist  for $m=1,m'=0$ at frequencies 
\begin{equation}
\label{eq:omegaps}
\omegap = \frac{\deltak\omega_\text{rf}\pm\Omega_c}{i-2j},
\end{equation}
with integers $j, \deltak$ such that $0\leq \deltak \leq i$ and $0\leq j < \lfloor i/2 \rfloor$. Here, $\deltak = |k_c'-k_c|$ corresponds to the difference between initial and final manifold and $i-2j = |\np'-\np|$ to the difference between initial and final probe photon number. The action of $\Vrf$ is confined to a single manifold such that $\Vp$ only couples states with $|k_c'-k_c|\leq 1$, resulting in the limit of $\deltak \leq i$. Thus only a finite number of transitions occur for a given order, as shown for first order in~\cite{Hofferberth2007}. 

Similarly, $|m'-m|\leq 1$, which means there is no first-order transition between states with $m = 1$ and $m' = -1$. Although these states couple in third order, the different paths interfere destructively, such that the total transition amplitude is zero and transitions do not occur.

\begin{figure}[h]
	\includegraphics[width = \textwidth]{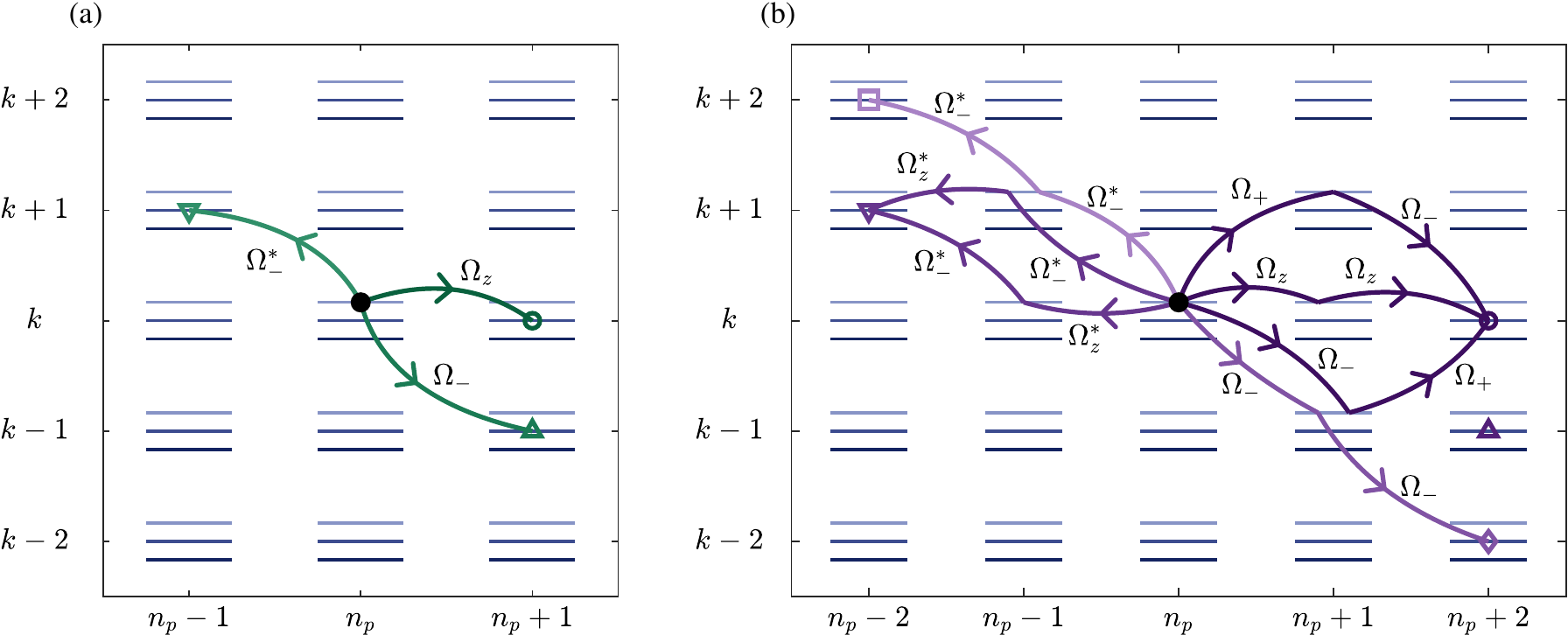}
	\caption{\label{fig:transitions}Paths of (a) first- and (b) second-order transitions of atoms dressed by a circularly-polarized field. The component of the probe field that drives the individual steps is indicated. The transition with frequency $(\omegarf+\Omega_c)/2$ is omitted from (b) for clarity, but proceeds mirrored to that at frequency $(\omegarf-\Omega_c)/2$, with the final state indicated by the upright triangle $\bigtriangleup$. \Tref{tab:transitions} shows transition frequencies and $\Omegaeff$ that correspond to the final states indicated here by the symbols.}
\end{figure}

\Tref{tab:transitions} shows transition frequencies and the corresponding $\Omegaeff$ for all first- and second-order transitions, as well as selected third-order transitions. The various paths contributing to first- and second-order transitions are shown in \fref{fig:transitions}. Note that while transitions at frequencies $2\omegarf\pm\Omega_c$ do not exist in first order, they do exist in third order and higher. The full transition amplitude for a given probe frequency is the sum over contributions from all orders. When the probe amplitude is weak compared to the dressing field amplitudes, however, the lowest order terms dominate this sum.

\begin{table}[h]
	\caption{\label{tab:transitions}Frequencies and matrix elements for all first- and second-order transitions with a circularly-polarized dressing field. For third order, we show only the transitions at $\Omega_c/3$ and $\omegarf\pm\Omega_c/3$ (and for $\Omega_c/3$, only the dominant terms proportional to $\Omegaz{}^3$). The markers in the first column correspond to those used in figures \ref{fig:transitions} and \ref{fig:transitionStrengthVSDetuning}.
	}
	\begin{indented}
		\item[]\begin{tabular}{@{}llll}
			\br
			& order & \textrm{$\omegap$} & \textrm{$\Omegaeff$}\\
			\mr
			\textcolor{Omega}{\fullmoon} & 1 & $\Omega_c$ & $\frac{\Omega_0}{\Omega_c}\Omegaz{}$ \\
			\bs
			\textcolor{omegaPOmega}{$\bigtriangleup$} & 1 & $\omegarf + \Omega_c$ & $\frac{1}{2}\frac{\Omega_c - \Delta}{\Omega_c}\Omegam{}$ \\
			\bs
			\textcolor{omegaMOmega}{$\bigtriangledown$} & 1 & $\omegarf - \Omega_c$ & $-\frac{1}{2}\frac{\Omega_c + \Delta}{\Omega_c}\Omegam{}^*$ \\
			\mr
			\bs
			\textcolor{OmegaH}{\fullmoon} & 2 & $\Omega_c/2$ & $\frac{2\Omega_0}{\Omega_c^2}\left(-\frac{\Delta}{\Omega_c}\Omegaz{}^2 - \frac{\Omega_c(2\omegarf+\Delta)}{8\omegarf^2-2\Omega_c^2}\Omegap{}\Omegam{}\right)$ \\
			\bs	
			\textcolor{HomegaPOmega}{$\bigtriangleup$} & 2 & $\frac{1}{2}(\omegarf+\Omega_c)$ &  $-\frac{1}{\Omega_c^2}\Omegam{}\Omegaz{}\left(\frac{\Omega_0^2}{\omegarf-\Omega_c}-\frac{(\Omega_c-\Delta)\Delta}{\omegarf+\Omega_c}\right)$ \\
			\bs
			\textcolor{HomegaMOmega}{$\bigtriangledown$} & 2 & $\frac{1}{2}(\omegarf-\Omega_c)$ &  $\frac{1}{\Omega_c^2}\Omegam{}^*\Omegaz{}^*\left(\frac{\Omega_0^2}{\omegarf+\Omega_c}+\frac{(\Omega_c+\Delta)\Delta}{\omegarf-\Omega_c}\right)$ \\
			\bs
			\textcolor{omegaPOmegaH}{$\Diamond$} & 2 & $\omegarf + \frac{1}{2}\Omega_c$ & $-\frac{1}{2}\frac{\Omega_0(\Omega_c - \Delta)}{\Omega_c^3}\Omegam{}^2$ \\ 
			\bs
			\textcolor{omegaMOmegaH}{$\Box$} & 2 &$\omegarf - \frac{1}{2}\Omega_c$ & $\frac{1}{2}\frac{\Omega_0(\Omega_c + \Delta)}{\Omega_c^3}\Omegam{}^{*2}$ \\
			\mr
			\bs
			& 3 & $\Omega_c/3$ & $\frac{9\Omega_0}{2^4\Omega_c^5}(8\Delta^2-\Omega_0^2)\Omegaz{}^3$ \\
			\bs
			& 3 & $\omegarf + \Omega_c/3$ & $\frac{3^4}{2^7}\frac{(\Omega_c-\Delta)\Omega_0^2}{\Omega_c^5}\Omegam{}^3$ \\
			\bs
			& 3 &  $\omegarf - \Omega_c/3$ & $-\frac{3^4}{2^7}\frac{(\Omega_c+\Delta)\Omega_0^2}{\Omega_c^5}\Omegam{}^{*3}$ \\
			\bs
			\br
		\end{tabular}
	\end{indented}
\end{table}

In~\cite{Perrin2017}, $i^{th}$-order transitions at frequencies $\Omega_c/i,\omegarf\pm\Omega_c/i$ were also predicted and their strengths calculated only for specific polarizations of the probe field, and by taking a second \ac{rwa}. In contrast, our method makes no such approximations, and our results differ accordingly: for arbitrary polarizations of the probe field, additional resonances appear (\eg{} with $\omegap = 1/2(\omegarf\pm\Omega_c)$). While our transition amplitudes agree with~\cite{Perrin2017} for first and second order, they differ for third order and higher.

\subsection{Single radiofrequency -- linear polarization}
\label{sec:linSRF}
The scenario of an atom dressed with a linearly-polarized field is more complicated, because the presence of counter-rotating terms causes the dressed eigenstates to contain bare states from an infinite number of manifolds. Thus, an infinite number of possible transitions exist at any order of the probe field, although most are of negligible strength. Ultimately, transitions  occur at the same  frequencies as for a circularly-polarized dressing field~\eref{eq:omegaps}, but many of these  occur at a lower order, and are thus stronger, when atoms are dressed by a linearly-polarized field.

Truncating the Hilbert space allows $H_1$ to be expressed as a finite-dimensional matrix  which can be diagonalized numerically to high accuracy. Equations~\eref{eq:energyconservation} and \eref{eq:norder} can then be solved analytically using the resulting eigenstates. The frequencies of the transitions are given by $\omegap = (\deltak\omega_\text{rf}\pm\Omega)/(i-2j)$, as for the circular case~\eref{eq:omegaps}, but with no limit on the integer $\deltak = k' - k$. $\Vp$ contains terms of any power in $b,b^\dagger$ as compared to \eref{eq:Vpt} which is linear in $b,b^\dagger$. This means that changes with $\deltak > 1$ are possible through absorption or emission of a single probe photon, resulting in an infinite number of transitions for any given order. 

Each operator $b$ or $b^\dagger$ introduces a factor to the effective Rabi frequency that is proportional to the dressing field amplitude. This results in dominant transitions at $\deltak = 0,1$ for a weak dressing field, but at larger values of $\deltak$ for a strong dressing field, as was shown in~\cite{Hofferberth2007}. First-order transitions at $\deltak\omegarf\pm\Omega$ are driven by the longitudinal component of the probe field for even~$\deltak$ and by the circularly-polarized component of the field for odd~$\deltak$.

In the limit of large frequency detuning, the order of the dressing field is well-defined~\cite{Garraway2016}. On resonance this is not the case, since the number of dressing photons for a given eigenstate is indeterminate. Instead, as an approximation, we use $\deltak$ as explained above.

\begin{figure}[h]
	\includegraphics[width = \textwidth]{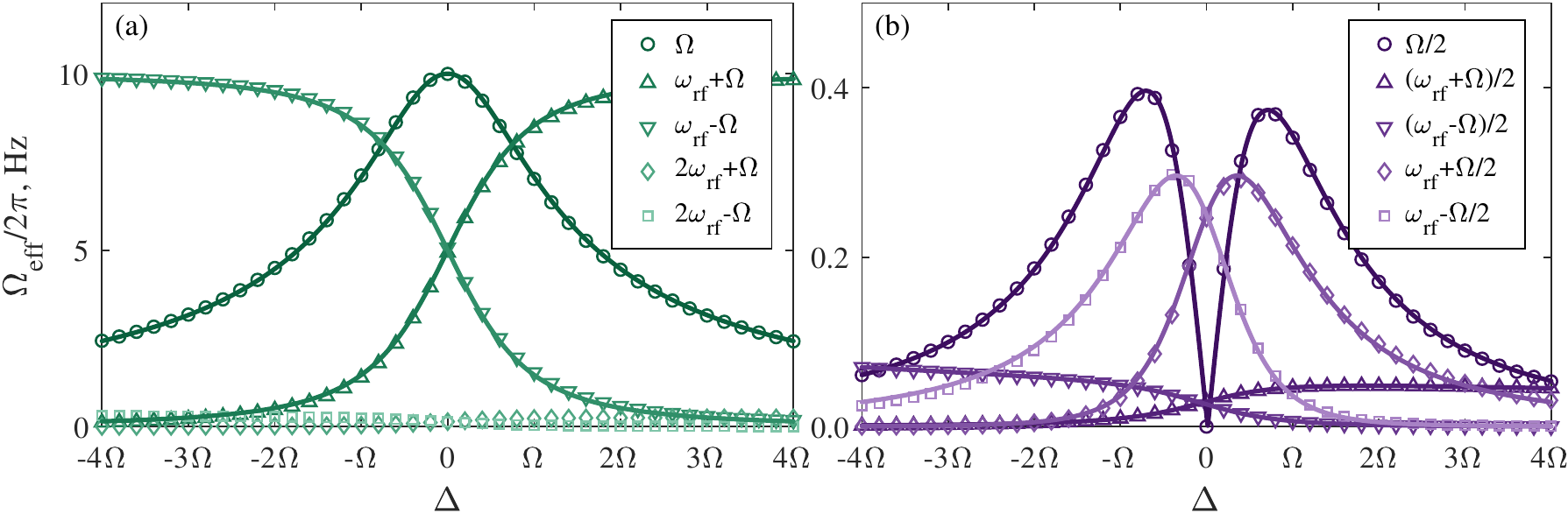}
	\caption{\label{fig:transitionStrengthVSDetuning}Amplitudes of (a) first- and (b) second-order transitions in a circularly-polarized (lines, analytical solutions) and a linearly-polarized (points, numerical solutions) dressing field with a frequency of \SI{3.6}{\MHz} and a Rabi frequency of \SI{0.2}{\MHz}, as a function of detuning $\Delta$. The amplitudes are calculated for a probe field on resonance, and $\Omegaz{} = \Omegap{} = \Omegam{} = \twopi\SI{10}{\kHz}$.}
\end{figure}

\Fref{fig:transitionStrengthVSDetuning} shows the dominant amplitudes for transitions that are first and second order in the probe field, as a function of $\Delta$, and for the cases of circularly- and linearly-polarized dressing \ac{rf} fields. In the limits $|\Delta|\rightarrow\infty$, the amplitudes for the various transitions approach those of the undressed Zeeman states, as expected~\cite{Garraway2016,Perrin2017}.

\subsection{Multiple radiofrequencies}
\label{sec:linMRF}
In this section we explain the spectrum expected when dressing with multiple dressing frequencies and a linearly-polarized field. For a dressing field containing multiple frequencies that are all multiples of a fundamental frequency $\omegaf$, the frequencies at which resonant transitions exist are given by a formula similar to \eref{eq:omegaps} but with $\omegarf$ replaced by $\omegaf$:
\begin{equation}
(\deltak\omegaf\pm\Omega)/(i-2j).
\end{equation}
Transitions are spaced by the fundamental frequency $\omegaf$, leading to a finely spaced comb of frequencies if $\omegaf$ is small. As for the case of a linearly-polarized dressing field with a single frequency, the expression for $\Vp$ in the case of multiple frequencies contains any combination of the operators  $b_i,b_j^\dagger$. The dominant contribution to any given transition is given by the minimum number of raising and lowering operators required as well as their respective prefactors\footnote{This is true unless several paths cancel each other, such that a different term dominates.}. For example, in a dressing field at frequency $\ki{1}\omegaf$, only one creation or annihilation is necessary to drive the transition for which $\deltak = \ki{1}$. For two dressing fields at frequencies $\ki{1}\omegaf, (\ki{1}+1)\omegaf$, the transition at $\deltak = 1$ takes two operators ($a_{\ki{1}+1}^\dagger a_{\ki{1}} = a_1^\dagger$), but the transition with $\deltak = \ki{1}/2$ takes $\ki{1}$ operators: $(a_{\ki{1}+1}^\dagger)^{\ki{1}/2}(a_{\ki{1}})^{\ki{1}/2} = a_{\ki{1}/2}$. Thus transitions at $(\ki{1}\omegaf\pm\Omega)/(i-2j), (\omegaf\pm\Omega)/(i-2j)$ are strong compared to transitions at $(\ki{1}/2\ \omegaf\pm\Omega)/(i-2j)$. This qualitative description explains the overall pattern of transition strengths in the experimental observations reported below and can lead to second-order transitions surrounding the frequency $(\ki{1}/2)\omegaf$ being stronger than first-order transitions at similar frequencies.

\section{Comparison of predicted and experimentally observed transitions}
\label{sec:comparison}
In the following, we compare the experimental results to the theoretical predictions. We calculate the transition frequencies and amplitudes for our experimental parameters up to fourth order for the \ac{srf} case and up to third order for the \ac{mrf} case, and compare these to the experimental spectra.
\subsection{Single radiofrequency}
\begin{figure}[h]
	\includegraphics[width = \textwidth]{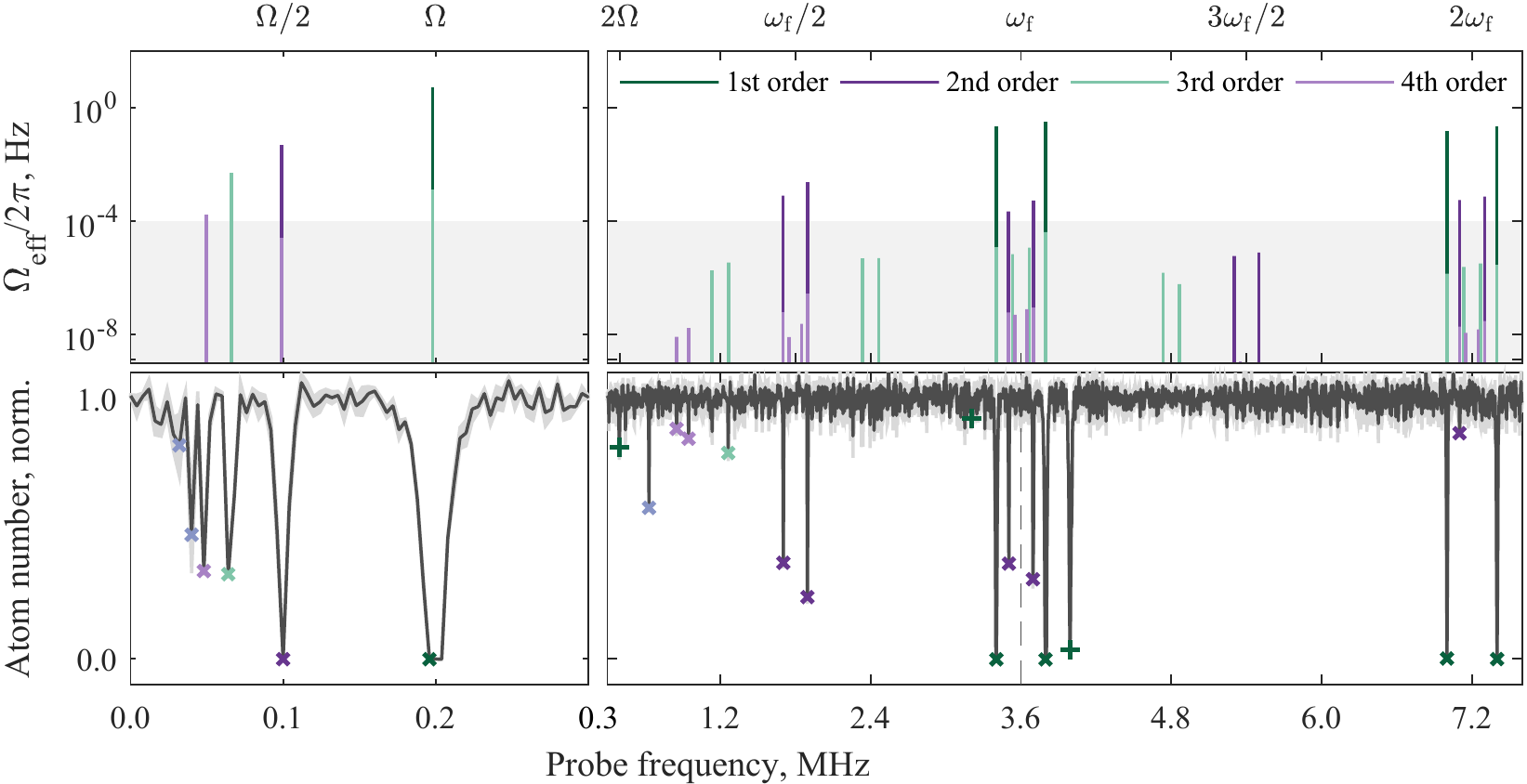}
	\caption{\label{fig:linSRF}The spectrum for atoms in a \ac{srf} dressing field. The predicted resonances (top) correspond to probe frequencies where loss is observed (bottom). Top: The magnitude of the matrix element is indicated by the length of the bar, with different colours indicating the order of the transition. Probe values are chosen as $\Omegaz{} = \twopi\SI{30}{\kHz},\Omega_\pm = \twopi\SI{1}{\kHz}$. The shading gives a rough indication of the threshold amplitude below which resonances cannot be detected, however there are exceptions, \eg{} the resonances near \SI{1.2}{\MHz}. The angular frequency of selected points is indicated on the upper $x$ axis. Bottom: The full experimental spectrum, as explained in \sref{sec:experiment}. Resonances identified from the data are marked with crosses, where the colour indicates the order. Blue markers indicate that the resonance is higher order than fourth. An `$\times$' indicates that the resonance was predicted in \sref{sec:linSRF} and a `+' that it arises because of the non-linear Zeeman effect (see \sref{sec:nonlinearZeeman}). The frequency of the dressing field is indicated by a vertical dashed line.}
\end{figure}

\Fref{fig:linSRF} shows the loss spectrum versus probe frequency for a linearly-polarized \ac{rf}-dressing field. The theoretical spectrum is displayed above the data.

For a circularly-polarized dressing field (data not shown), we observe the same transitions, but the amplitude for those at $2\omegarf\pm\Omega$ is much reduced. The remaining peaks are too strong to result purely from third-order effects, and we attribute their observed strength to imperfections in the circular polarization.

To calculate theoretical values for $\Omegaeff$, the value for $\Omegaz{}$ was taken to be $\twopi\SI{30}{\kHz}$ as estimated in \sref{sec:experiment} and the values for $\Omegap{} = \Omegam{} = \twopi\SI{1}{\kHz}$ were chosen such that the theoretical spectrum replicates the experimental one. As expected from the geometry of the coil array, these values are significantly lower than $\Omegaz{}$. All first-order coupling strengths are of a similar order of magnitude due to the following coincidence of our apparatus: transitions at low frequencies mainly couple via $\Omega_z$, and transitions at higher frequencies mainly couple via $\Omega_-$; $\Omega_z \gg \Omega_-$, but the amplitude of the probe field drops by approximately \SI{30}{\decibel} below \SI{0.7}{\MHz}. 

Since the experimental values for polarization and amplitude of the probe field are only approximate, a more quantitative comparison cannot be made. The grey shaded area approximately indicates which transitions are visible. Our detection of some transitions with strengths below this threshold suggests that some transition amplitudes are underestimated, particularly at lower frequencies.

We identify experimentally observed resonances with theoretically predicted ones if the frequencies agree within the resolution of the applied probe frequencies, that is \SI{4}{\kHz} for the \ac{srf} case. \Fref{fig:linSRF} shows resonances at frequencies of \SIlist[list-units=single]{0.196; 3.404; 3.796; 7.0; 7.4}{\MHz}, which we identify with all first-order resonances predicted by \eref{eq:omegaps} in the range \SIrange[range-units=single]{0}{7.5}{\MHz}, that is for $\deltak = 0,1,2$. Theoretically predicted transition strengths are on the order of $\Omegaeff = \twopi\SI{1}{\Hz}$ and all atoms are lost rapidly. 

Second-order resonances can be identified with those predicted for $\deltak = 0, 1, 2$ at frequencies \SIlist[list-units=single]{0.1; 1.704; 1.896; 3.504; 3.696}{\MHz} as well as at \SI{7.1}{\MHz} for $\deltak = 4$. The dominant contribution to the transition strength for $\deltak = 3$ is proportional to $\Omegaz{}\Omegap{}$, which makes it weaker than transitions for $\deltak = 4$ which are proportional to $\Omegaz{}^2$. Second-order Rabi frequencies are on the order of $\Omegaeff = \twopi\SI{e-3}{\Hz}$, and slightly more than half the atoms are lost within $\tp = \SI{1.2}{\sec}$.

Furthermore, we observe resonances at \SIlist[list-units=single]{64; 48; 40; 32}{\kHz} which correspond to $i^{th}$ order resonances at $\Omega/i$ for $3\leq i\leq6$. Although these are predicted to be weaker than some transitions we do not observe, third and fourth-order resonances also appear at frequencies \SIlist[list-units=single]{1.264; 0.852; 0.948}{\MHz}, \ie{} for $\omegap = (\omegarf+\Omega)/3, (\omegarf\pm\Omega)/4$. We observe one sixth-order resonance at a frequency of $\omegap = (\omegarf + \Omega)/6 = \SI{633}{\kHz}$. 
The FWHM of this transition is \SI{1}{\kHz} and we have verified that it is indeed sixth order by varying the Rabi frequency and observing the expected shift in the transition frequency. Only the first-order resonances and the resonance at $\omegarf+\Omega/2$ have been observed in previous work~\cite{Hofferberth2007,Perrin2017}. 

Resonances at frequencies of \SIlist[list-units=single]{0.396; 3.208; 3.992}{\MHz} are visible, but do not correspond to any of the transitions predicted in \sref{sec:linSRF}. They can be matched to transitions at $\deltak\omegarf\pm2\Omega$ for $\deltak = 0, 1$, frequencies which correspond to the energy difference between states with $m = 1$ and $m' = -1$. In \sref{sec:theoryResults} we showed that transitions between these states do not exist under our assumptions. In the following section we argue that the non-linearity of the Zeeman effect needs to be taken into account in order to explain the appearance of these resonances.

\subsection{Non-linear Zeeman effect}
\label{sec:nonlinearZeeman}
The eigenenergies of an atom in a static magnetic field can be calculated from the Breit-Rabi formula, and the influence of deviations from linearity on dressed atoms has been investigated in~\cite{Sinuco-Leon2012}. The static magnetic field used in this work causes Zeeman splittings of a few \SI{}{\MHz}, compared to the hyperfine splitting of \SI{6.8}{\GHz}. Therefore the energies of the undressed Zeeman states can usually be calculated with sufficient precision without taking into account non-linearity. However, the asymmetry of the eigenenergies does have an observable influence on  transitions that are `forbidden' in the linear regime, analogous to a breakdown of the selection rules. We incorporate it into our calculations by introducing an additional term so that the Hamiltonian now reads
\begin{equation}
H_0 \rightarrow H_0 + \hbar\delta_\text{ZM}\id\otimes\ket{-1}\bra{-1},
\end{equation}
where $\ket{-1}$ denotes the Zeeman state with $m_F = -1$ and $\hbar\delta_\text{ZM}$ is the difference between the energy separation of the $m_F = -1,0$ and the $m_F = 0,1$ states, such that $\delta_\text{ZM} = 0$ in the linear Zeeman effect. For the low field strengths investigated here, $m_F$ is still a good quantum number and $\ket{-1}$ is still a well-defined state.

We calculate the eigenstates of $H_1 = H_0 + \Vrf$ numerically and find first-order transition frequencies and amplitudes as described above. A significant feature is that transitions at probe frequencies of $\deltak\omegaf\pm2\Omega$ now exist between states with $m = 1$ and $m' = -1$, whereas none are predicted for a purely linear Zeeman effect, as in~\cite{Hofferberth2007}. These transitions are stronger at larger static fields where the non-linear Zeeman shift is increased.

We test this conjecture by probing the resonance $\omegap = 2\Omega$ experimentally, for different values of $\omegarf$, thus varying $\omega_0$ and the magnetic field at the position of the atoms. We observe increased loss as the magnetic field and thus the non-linearity of the Zeeman splitting increases, as expected.

\subsection{Multiple radiofrequencies}
\label{sec:MRFresults}
With multiple dressing frequencies present, we expect transitions to repeat at the common fundamental frequency, as detailed in \sref{sec:linMRF}. The transition strength not only depends on the probe polarization and amplitude, but also on the integer $\deltak$ in \eref{eq:omegaps}. 

\begin{figure}[h]
	\includegraphics[width = \textwidth]{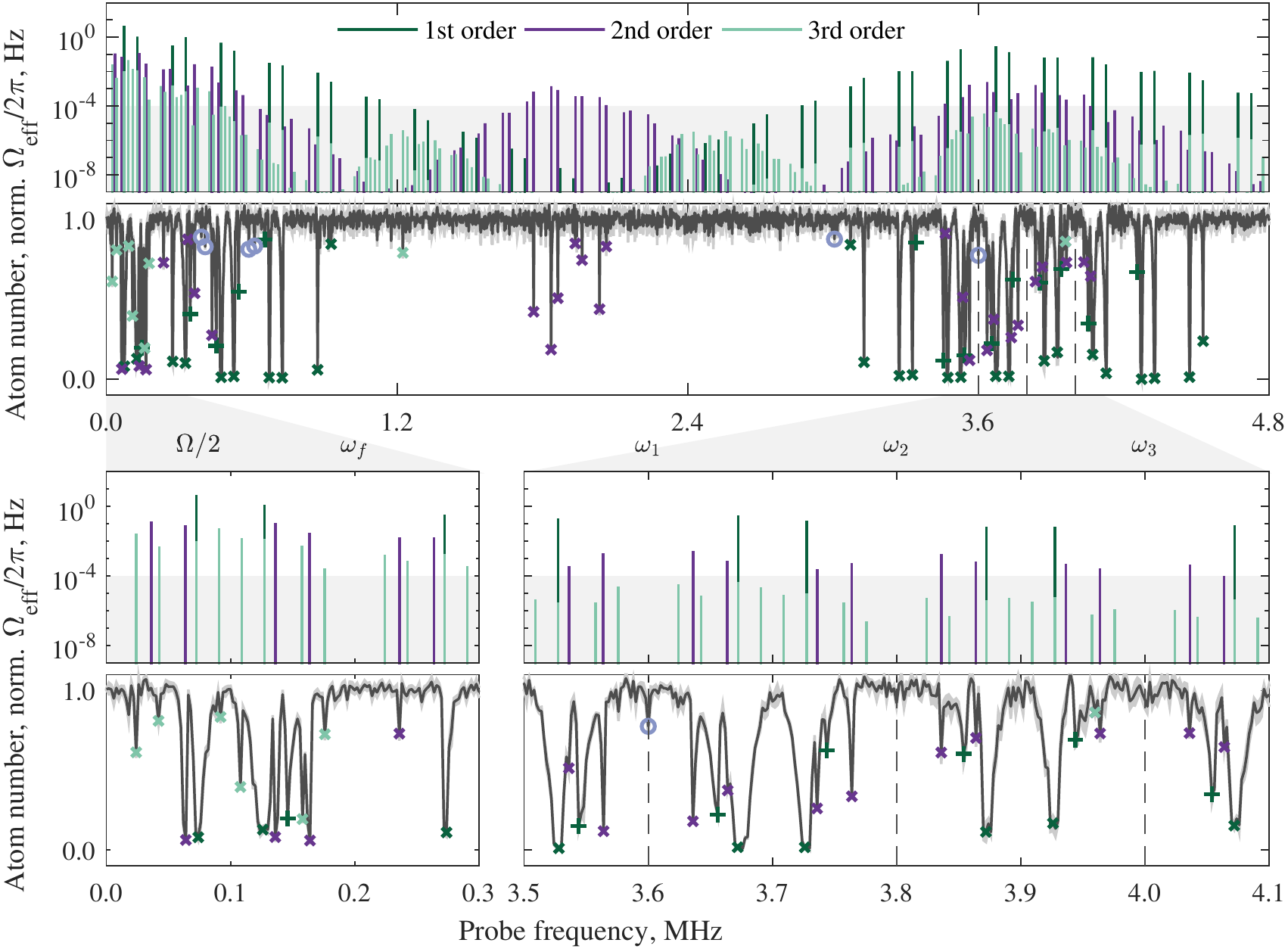}
	\caption{\label{fig:linMRF}Normalized atom number versus probe frequency and theoretically predicted transitions for the linear \ac{mrf} case. The plot style is consistent with \fref{fig:linSRF}. Unidentified resonances are indicated by blue circles, and ranges \SIrange{0}{0.3}{\MHz} and \SIrange{3.5}{4.1}{\MHz} are enlarged.}
\end{figure}

\Fref{fig:linMRF} shows the experimental spectrum with theoretical predictions displayed above. In the theoretical spectrum, a periodicity of transition amplitudes of order $i$ with $\omega_1/i$ is evident. As explained in \sref{sec:linMRF}, this is due to the fact that approximately $\ki{1}/2 \approx 10$ creation or annihilation operators are required for transitions with $\deltak = \ki{1}/2$. In contrast, the second-order transitions at frequency $(\ki{1}\omegaf\pm\Omega)/2$, require one creation operator only, since $\deltak = \ki{1}$. This results in second-order transitions being significantly stronger than first-order ones surrounding the probe frequencies near $\omega_1/2$. The same behaviour can be observed for third-order transitions. The periodicity with $\omega_1$ is a result of atoms being trapped in the potential well that is near-resonant with $\omega_1$. We note that for larger probe frequencies, the periodic behaviour dephases and becomes washed out because of contributions from $\omega_{2,3}$.

These predictions agree well with the observations: most of the visible resonances are at low frequencies, as well as surrounding the dressing frequencies. Additionally, some transitions at $(\omegarf\pm\Omega)/i$ are visible for $i = 2, 3$ and for $\omegarf = \omega_1,\omega_2,\omega_3$. Again, transitions at $\deltak\omegaf\pm2\Omega$ appear, as a result of the non-linearity of the Zeeman effect.

We observe atom loss at probe frequencies close to multiples of the fundamental frequency, as indicated by blue circles in \fref{fig:linMRF}, which cannot be reconciled with theoretically predicted resonances. We attribute this to a sudden perturbation in the dressing potential when the probe field is turned on at these frequencies, increasing the amplitude of the dressing field suddenly. The resulting kick imparted to the atoms causes heating and loss. Given that the amplitudes of the dressing fields are only about three times stronger than that of the probe field, this is unsurprising.

\section{Conclusion and outlook}
We have presented a general theoretical framework to calculate transition frequencies and strengths for atoms dressed by multiple frequencies. Accounting for arbitrary polarizations and higher-order transitions provides a complete description of the spectrum. This methodology enabled us to derive analytical expressions for transition frequencies and amplitudes for the case of a single, circularly-polarized field. Numerical diagonalization was used for cases beyond the \ac{rwa} and for multiple-frequency dressing fields. Transition frequencies for any order in the probe field are given by a simple formula. Our results are consistent with previous predictions~\cite{GarridoAlzar2006,Hofferberth2007,Perrin2017}, but greatly extend these. We uncover transitions that were not predicted previously by considering an arbitrary polarization of the probe and dressing fields as well as calculating transitions of higher order in the probe field. Spontaneous emission was neglected, but could be included in our model if required, enabling investigation of an even wider range of multi-frequency phenomena.

We performed \ac{rf} spectroscopy of atoms trapped in single- and multi-frequency fields, observing transitions up to sixth order in the probe field. Transitions between states with $m = 1, m' = -1$ were observed, which were not apparent in previous work; these arise from the non-linearity of the Zeeman effect. This effect introduces an asymmetry into the Hamiltonian operator that results in a coupling of states with $m = 1, m' = -1$, thus causing a breakdown of the selection rules.

The observed atom loss agrees qualitatively with the predicted transition strengths, though we also observe loss for surprisingly low effective Rabi frequencies. Several details of the experiment prevent a more detailed quantitative comparison: firstly, the polarization and amplitude of the probe field is undetermined. Secondly, the \SI{4}{\kHz} spacing of data points means that we do not accurately determine the maximum loss rate, especially for the intrinsically narrower higher-order transitions. A clear example of this lack of resolution occurs for the observed sixth-order resonance. A measurement with finer spacing over a short range determined the FWHM to be \SI{1}{\kHz}, but it would be both impractical and unnecessary to apply this resolution over the whole range of frequencies inspected here. It is likely that some high-order resonances were missed by our experimental procedure, but more than a sufficient number of transitions were observed to give an extremely high degree of confidence in the theoretical model, and all observed transitions in the \ac{srf} case are explained. We observe higher-order transitions even for moderate probe amplitudes, highlighting the importance of taking these into account when determining the effect of stray fields.

The description of multi-frequency fields is common in non-linear optics, and it is insightful to compare this approach to our methods: in non-linear optics the amplitude of dipole oscillations is typically small, off-resonant, and at frequencies associated with the driving fields. In contrast, the multi-frequency transitions considered here can have a large amplitude but are typically slowly oscillating and are observed on resonance. This work is therefore at the opposite end of the scale of multiple-frequency effects to the standard perturbative approach in non-linear optics. These extremes can be combined with the more general theoretical methods presented in~\cite{Yuen2018}.

Although the theoretical framework presented here is widely applicable, we have focussed on the experimental implementation of \ac{mrf} dressed potentials.
This powerful tool for confining cold atoms increases the versatility of magnetic trapping techniques. In our previous work, we employed this technique to observe matter-wave interference~\cite{Bentine2018}. Extending the use of these potentials, \eg{} to investigate thermalization in two-dimensional gases~\cite{Mathey2010}, requires a reduction of the well-spacing. For this, a more detailed understanding of the plethora of resonances that arises when multiple dressing fields are present is essential. In particular, our work provides a framework to understand and mitigate the loss of atoms through spurious noise when working with \ac{mrf} dressed potentials. This is of critical importance for experiments that use many closely-spaced frequencies, such as the proposal to form a periodic potential of individually controllable wells~\cite{Courteille2006}.

\section*{Acknowledgements}

The authors would like to thank Jordi Mur-Petit for useful discussions and David Garrick for comments on the manuscript. This work was supported by the  EU  H2020  Collaborative  project  QuProCS  (Grant Agreement 641277) and by the EPSRC grant (Reference EP/S013105/1). KL, EB, and AJB thank the EPSRC for doctoral training funding. The authors acknowledge thoughtful input from the referee that has improved the manuscript.

\appendix
%\section{Choice of states to describe the dressing fields}
\section{Description of the dressing field using non-degenerate states}
%\section{The set of non-degenerate states used to describe the dressing field}
\label{sec:basis}
For $l$ dressing fields with frequencies $\ki{q}\omegaf$, we can describe the state of the dressing field in the Fock basis $\left\lbrace\ket{N_1,N_2,\dots,N_l}\right\rbrace$. This basis is degenerate, however, and the interaction $\Vrf$ connects degenerate states in higher orders. For the case of three dressing fields with $\ki{1}, \ki{2}, \ki{3} = 1,2,3$, the two states $\ket{N_1,N_2,N_3,1}$ and $\ket{N_1+1,N_2-2,N_3+1,1}$ are degenerate and connected via $a_2 F_+ a_1^\dagger F_- a_2 F_+ a_3^\dagger F_-$. 

No matter how large the basis is made, any truncation excludes states which are similarly degenerate. Numerical diagonalization of the resulting matrix is problematic and leads to erroneous avoided crossings as well as states with different energies where they should be degenerate~\cite{Ben}.

To avoid this problem, we represent the coherent state of our dressing fields $\ket{\alpha_{1},\alpha_{2},\dots,\alpha_{l}}$ 
% using a subspace of the full Hilbert space.
using an orthonormal set of non-degenerate states $\lbrace\ket{N}\rbrace$. This set is the basis of a subspace containing the states of the Hilbert space that are relevant to our calculations.
%The basis for this subspace consists of an orthonormal set of non-degenerate states, $\lbrace\ket{N}\rbrace$. 
Each state $\ket{N}$ is proportional to the projection of $\ket{\alpha_{1},\alpha_{2},\dots,\alpha_{l}}$ onto the set of states with energy $N \hbar \omega_f$ and is normalised to one. Thus, in the Fock basis $\ket{N}$ is a superposition of all Fock states with energy $\sum_q N_q \ki{q} \hbar \omegaf = N \hbar \omegaf$. The amplitude of each Fock state is proportional to its amplitude within $\ket{\alpha_{1},\alpha_{2},\dots,\alpha_{l}}$, so explicitly depends on the set of $\alpha_q$ describing the coherent dressing field. Hence we can faithfully write $\ket{\alpha_{1},\alpha_{2},\dots,\alpha_{l}}$ as the superposition $\sum_N \gamma_N \ket{N}$ with amplitudes $\gamma_N$. For a single frequency field, the set $\lbrace \ket{N} \rbrace$ corresponds to the standard Fock basis. 

The action of the creation and annihilation operators on each state in the subspace basis $\lbrace \ket{N} \rbrace$ is $a_q \ket{N} = \alpha_q[1-\varepsilon(N)] \ket{N-n_q}$ and $a_q^{\dagger} \ket{N} = \alpha_q^*[1-\varepsilon(N)]\ket{N+n_q}$, as shown in ~\cite{Ben}. $\varepsilon(N)$ is a small, $N$-dependent number that arises from quantum fluctuations. To leading order $\varepsilon \sim (N-\langle N \rangle) / \langle N \rangle$ and is much smaller than unity when the expectation value $\langle N \rangle \gg 1$. Thus to a very good approximation $(a_q/\alpha_q) \ket{N} = \ket{N-n_q}$ and $(a_q^{\dagger}/\alpha_q^*) \ket{N} = \ket{N+n_q}$, where the terms in brackets are the normalized operators used in \eref{eq:normaloperators}. Thus the set of states $\lbrace\ket{N}\rbrace$ is closed under the action of $a_q$, and for $a_q^{\dagger}$ except for a small renormalisation which is negligible when $\abs{\alpha_q}^2 \gg 1$~\cite{Ben}, as is the case for our experiment. Therefore, all possible final states are also contained within this subspace, allowing us to calculate the matrix elements for all \ac{rf} transitions for an atom dressed by the strong coherent fields used in this work.

Using a common fundamental frequency in this derivation may seem like a limitation. It requires the individual frequencies to be rational, but since the rational numbers are dense in the real numbers, this does not pose a problem in reality.

\section{Labelling the dressed eigenstates}
\label{sec:manifolds}
\begin{figure}[h]
	\includegraphics[width = \textwidth]{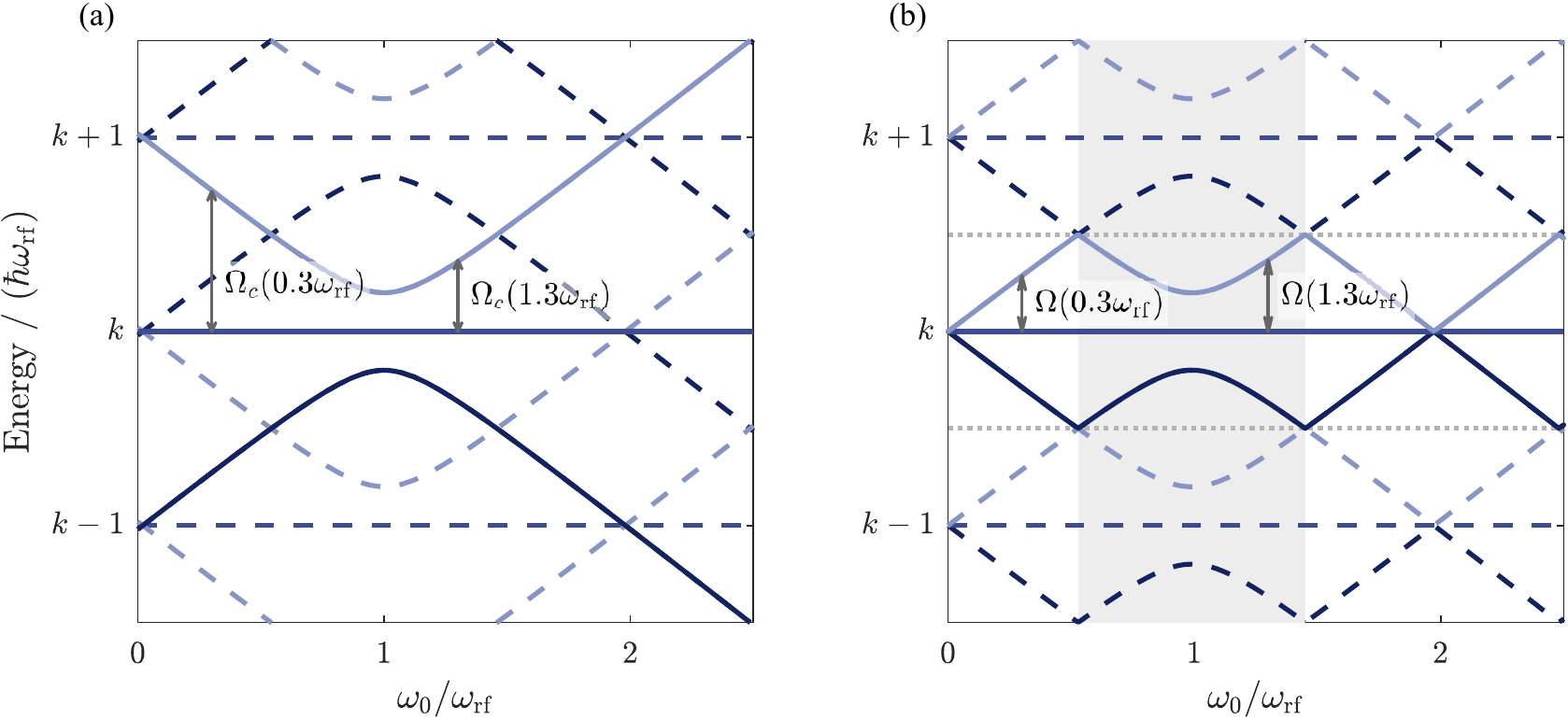}
	\caption{\label{fig:manifolds}Eigenenergies of an atom dressed by a single-frequency field. States with labels $\ketprime{k,1}$ are light blue, and states with $\ketprime{k,-1}$ are dark blue. States with a specific value of $k$ are shown as solid lines, and other states as dashed lines. (a) For a circularly-polarised dressing field, $k_c = N-m_F$ is a conserved quantity, and the generalized Rabi frequency $\Omega_c = (\Omega_0^2+\Delta_2)^{1/2}$ can be found analytically. (b) Using the definition of \eref{eq:eigenenergies} is a convenient way to label eigenstates for arbitrary dressing fields. Within the vertical shaded area, the two definitions coincide for the parameters shown here. (c) The basis definition of \eref{eq:eigenenergies} illustrated for a multiple-frequency dressing field with frequency components \SIlist[list-units=single]{3.6; 3.8; 4.0}{\MHz}.}
\end{figure}

A circularly-polarised single-frequency dressing field couples the states within a manifold of constant $k_c = N-m_F$, and it is convenient to label the eigenstates by this quantum number as shown in \fref{fig:manifolds}~(a). The corresponding eigenenergies are $\hbar (k_c\omegarf + m \Omega_c)$, where $m$ labels the states within the manifold and $\Omega_c$ is the generalized Rabi frequency.

For systems where this manifold number is not conserved under the application of $\Vrf$, it does not constitute a good label for the eigenstates of $H_1$. This is true even for a linearly-polarised field, but typically the \ac{rwa} is taken, under which $k_c$ is conserved. For multiple dressing fields, these quantum numbers are unwieldy. Instead, we define our quantum numbers as illustrated in \fref{fig:manifolds}~(b), noting that the eigenenergies are periodic with $\omegaf$. Our state labels are defined locally to each value of $\omega_0$, with corresponding eigenenergies $\hbar(k\omegaf + m \Omega(\omega_0))$. The values $m$ label the states that are grouped by the same value of $k$, with $\Omega$ the energy difference between neighbouring states of equal $k$. Note that the definition that neighbouring states are separated by the energy $\hbar\Omega$ requires that states with different values of $k$ cross for $F > 1$.

The two definitions coincide for small dressing field amplitude and small detuning, as indicated in \fref{fig:manifolds}~(b) by the shaded area.

\section{Resolvent formalism}
\label{sec:resolvent}
Employing the resolvent formalism enables understanding of the processes on the level of quantum states, \eg{} to identify interference effects. It also provides the possibility of treating the probe field quantum mechanically. We derive effective Hamiltonians between pairs of resonant states using the resolvent formalism, details of which can be found in~\cite{Cohen-Tannoudji1992}.

By defining the resolvent operator $G(z) = 1/(z-H)$, algebraic rather than integral equations can be used to describe the time-evolution of $H$. The time-evolution operator can be retrieved by a contour integral of $G(z)$. We identify a subspace $\subspace$ that contains states which are important in the process that is investigated -- in our case these are initial and final states as well as all states which are close in energy. Projection operators $P, Q = \id - P$ project onto $\subspace$ and onto its complement. $G(z)$ projected onto $\subspace$ can then be rewritten as 
\begin{equation}
PG(z)P = 1/(z-PH_1P-PR(z)P) 
\end{equation}
with the level-shift operator $R(z)$. One can identify an effective Hamiltonian acting on $\subspace$ from this version of the resolvent: $H_\text{eff} = PH_1P + PR(z)P$. The level-shift operator can be written as a power expansion in $\Vp$:
\begin{equation}
R(z) = \Vp + \Vp\frac{Q}{z-H_1}\Vp + \Vp\frac{Q}{z-H_1}\Vp\frac{Q}{z-H_1}\Vp + \dots
\end{equation}
A common approximation is to replace $z$ with $E_0$, the mean energy of states in $\subspace$. This is valid provided the energy shift due to $\Vp$ is small compared to the energy difference of intermediate states. In our case of assuming the probe field to be weak in comparison to the dressing fields, this approximation is valid.

The level-shift operator describes interactions between two states in $\subspace$ via intermediate states in the complement. The terms $Q/(E_0-H_1)$ are propagators in frequency space. Truncating the series thus results in a cut-off in energy space, as compared to time-evolution operators in time-dependent perturbation theory, where truncation results in a cut-off in time. The $i^{th}$ term in the expansion of $R(z)$ corresponds to a path via $i-1$ intermediate states and matrix elements of this term describe transitions of $i^{th}$ order in the probe field.

\section*{References}
\bibliographystyle{iopart-num}
\bibliography{refs}

\end{document}